\documentclass[11pt]{article}
\usepackage{amssymb}
\usepackage{epsfig}
\usepackage{graphicx}
\usepackage[nomarkers]{endfloat}
\usepackage{pstricks}

\newcommand{\BE}{\begin{equation}}
\newcommand{\EE}{\end{equation}}
\begin{document}
\begin{titlepage}

\vspace*{1mm}
\begin{center}

   {\LARGE{\bf Flat-space picture of gravity vs. General Relativity: a precision test
   for present ether-drift experiments}}

\vspace*{14mm}
{\Large  M. Consoli and E. Costanzo}
\vspace*{4mm}\\
{\large
Istituto Nazionale di Fisica Nucleare, Sezione di Catania \\
Dipartimento di Fisica e Astronomia dell' Universit\`a di Catania \\
Via Santa Sofia 64, 95123 Catania, Italy \\ }
\end{center}
\begin{center}
{\bf Abstract}
\end{center}
Modern ether-drift experiments in vacuum could in principle detect
the tiny refractive index that, in a flat-space picture of gravity,
is appropriate for an apparatus placed on the Earth's surface. In
this picture, in fact, if there were a preferred reference frame,
light on the Earth would exhibit a slight anisotropy with definite
quantitative differences from General Relativity. By re-analyzing
the data published by two modern experiments with rotating optical
resonators, and concentrating on the part of the signal that should
be free of spurious systematic effects, we have found evidences that
would support the flat-space scenario.

\end{titlepage}

\section{Introduction}

The present generation of ether-drift experiments, using rotating
optical resonators, is currently pushing the relative accuracy of
the measured frequency shifts to the level ${\cal O}(10^{-16})$. As
we shall try to illustrate, this level of accuracy is crucial to
obtain fundamental informations on the space-time structure of the
vacuum and the possible existence of a preferred reference frame.

In this paper, we'll present a re-analysis of the observations
reported in Ref.\cite{joint} for the anisotropy of the speed of
light in the vacuum. This re-analysis, that improves with a higher
statistics on our previous work \cite{previous} based on the data of
Refs.\cite{peters,schiller}, leads to two conclusions: i) the
experiments exhibit a non-zero daily average for the amplitude of
the signal ii) the magnitude of this average amplitude is entirely
consistent with the theoretical anisotropy parameter
\cite{pagano,pla,cimento} \BE \label{first} |B_{\rm th}|\sim 3({\cal
N}_{\rm vacuum}-1)\sim 42\cdot 10^{-10}\EE that, in the presence of
a preferred frame, enters the two-way speed of light in the vacuum
to ${\cal O}(v^2/c^2)$ \BE
{{\bar{c}(\theta)-\bar{c}(0)}\over{c}}\sim B
~{{v^2}\over{c^2}}\sin^2\theta \EE In Eq.(\ref{first}) ${\cal
N}_{\rm vacuum}$  indicates the effective vacuum refractive index
that is appropriate, in a flat-space picture of gravity, for an
apparatus placed on the Earth's surface.

We emphasize that the prediction Eq.(\ref{first}) refers to
experiments performed in the vacuum, i.e. in the highest vacua
attainable with present technology. Therefore, in principle, it
should not be compared with other type of experiments as those of
Ref.\cite{tobar2} where about $100\%$ of the electromagnetic energy
propagates within a dielectric medium with ${\cal N}\sim 3$
\cite{tobar1}.

After this general Introduction, the plan of the paper is as
follows. In Sect.2, we shall first illustrate the basic formalism
and report the experimental data of Ref. \cite{joint}. Then, in
Sect.3, we shall use these published data to deduce a basic quantity
of any ether-drift experiment: the daily average amplitude of the
signal $A_0$ that, so far, has never been reported by the
experimental groups. Further, in Sect.4, we shall try to estimate
the possible systematic uncertainty on our value of $A_0$, by also
comparing with Ref.\cite{schiller}. Then, in Sect.5, we shall point
out that the experimental values of $A_0$ for the two experiments,
besides being well consistent with each other, are also in good
agreement with the theoretical prediction expected in a flat-space
picture of gravity, in the presence of a preferred reference frame.
Finally, in Sect.6, we shall present our summary and conclusions.

\section{Basic formalism and experimental data}

The experimental data reported in Ref.\cite{joint} refer to 27
short-period observations, performed by the Berlin group
\cite{peters} during a total period of 392 days. The starting point
for our analysis is the expression for the relative frequency shift
of two optical resonators at a given time $t$. For the Berlin
experiment \cite{peters}, this can be expressed as \BE
\label{basic2}
      {{\Delta \nu (t)}\over{\nu_0}} =
      {S}(t)\sin 2\omega_{\rm rot}t +
      {C}(t)\cos 2\omega_{\rm rot}t
\EE
where $\omega_{\rm rot}$ is the rotation frequency of one resonator
with respect to the other which is kept fixed in the laboratory and
oriented north-south. We observe that the sine component of the
signal was denoted as $S(t)$ in Ref.\cite{peters} and as $B(t)$ in
Ref.\cite{joint}. Here, we shall maintain the notation of
Ref.\cite{peters} where
 the Fourier expansions of $S(t)$
and $C(t)$ are expressed as  \BE \label{amorse1}
      {S}(t) = S_0 +
      {S}_{s1}\sin\tau +{S}_{c1} \cos\tau
       + {S}_{s2}\sin(2\tau) +{S}_{c2} \cos(2\tau)
\EE
\BE
\label{amorse2}
      {C}(t) = {C}_0 +
      {C}_{s1}\sin\tau +{C}_{c1} \cos\tau
       + {C}_{s2}\sin(2\tau) +{C}_{c2} \cos(2\tau)
\EE
$\tau=\omega_{\rm sid}t$ being the sidereal time of the observation
in degrees and $\omega_{\rm sid}\sim {{2\pi}\over{23^{h}56'}}$.
Introducing the colatitude of the laboratory $\chi$, and the unknown
average velocity, right ascension and declination of the cosmic
motion with respect to a hypothetical preferred frame (respectively
$V$, $\alpha$ and $\gamma$), one finds the expressions reported in
Table I of Ref.~\cite{peters} in the RMS formalism \cite{rms} \BE
\label{C0}
      {C}_0 =-
      {{K \sin^2\chi}\over{8}} (3 \cos 2{\gamma} -1)
\EE
\BE \label{CS1}
      {C}_{s1}= {{1}\over{4}}K
      \sin 2{\gamma} \sin{\alpha} \sin 2\chi
      ~~~~~~~~~~~~~~~~~~~~~~~~
     {C}_{c1}={{1}\over{4}}K \sin 2{\gamma}
      \cos{\alpha} \sin 2\chi
\EE
\BE \label{CS2}
      {C}_{s2} = {{1}\over{4}}K \cos^2{\gamma}
      \sin2{\alpha}  (1+ \cos^2\chi)
~~~~~~~~~~~~
      {C}_{c2} = {{1}\over{4}}K \cos^2{\gamma}
      \cos2{\alpha} (1+ \cos^2\chi)
\EE
where \BE \label{cappa}
 K=|B|{{V^2}\over{c^2}}
\EE is proportional to the light anisotropy parameter. The
corresponding $S-$quantities are also given by ($S_0=0$) \BE
\label{s1}{S}_{s1}=-{{ {C}_{c1} } \over {\cos\chi}}~~~~~~~
{S}_{c1}={ {{C}_{s1} }\over{ \cos\chi\ }} \EE \BE \label{s2}
{S}_{s2}= -{{2\cos\chi}\over{1+\cos^2\chi}}{C}_{c2}~~~~~~~{S}_{c2}=
{{2\cos\chi}\over{1+\cos^2\chi}}{C}_{s2}\EE

\begin{table}
\caption{The experimental $C_k$ coefficients as extracted from Fig.2
of Ref.\cite{joint}. }
\begin{center}
\begin{tabular}{llll}
\hline\noalign{\smallskip} $C_{s1}[{\rm x}10^{-16}]$ & $C_{c1}[{\rm
x}10^{-16}]$ & $C_{s2}[{\rm x}10^{-16}]$ &
$C_{c2}[{\rm x}10^{-16}]$   \\
\noalign{\smallskip}\hline\noalign{\smallskip} \hline
$9.2\pm4.6$ &  $10.0\pm 5.4$ &  $-1.5\pm4.6$ &  $0.0\pm 5.4$ \\
$-2.3\pm6.9$ &  $-4.6\pm 6.2$ &  $-5.8\pm6.5$ &  $-0.8\pm 6.2$ \\
$-2.3\pm3.8$ &  $0.0\pm 3.1$ &  $-2.3\pm3.8$ &  $-4.6\pm 3.8$ \\
$4.6\pm4.6$ &  $11.5\pm 3.8$ &  $-14.6\pm3.8$ &  $-0.8\pm 4.6$ \\
$5.4\pm8.5$ &  $-11.5\pm 8.5$ &  $11.5\pm8.5$ &  $-4.6\pm 6.9$ \\
$-5.4\pm5.4$ &  $-6.2\pm 6.2$ &  $6.9\pm5.4$ &  $1.5\pm 5.4$ \\
$-10.8\pm6.2$ &  $-5.4\pm 5.4$ &  $-6.2\pm6.2$ &  $-4.6\pm 5.4$ \\
$-3.1\pm9.2$ &  $7.7\pm 9.2$ &  $-10.8\pm9.2$ &  $7.7\pm 8.5$ \\
$-6.2\pm6.2$ &  $6.2\pm 7.7$ &  $-2.7\pm6.5$ &  $6.9\pm 6.2$ \\
$-1.5\pm4.6$ &  $-9.2\pm 4.6$ &  $-10.0\pm5.4$ &  $-3.1\pm 5.4$ \\
$2.3\pm3.8$ &  $4.6\pm 3.1$ &  $-3.1\pm3.1$ &  $0.0\pm 3.8$ \\
$13.8\pm7.7$ &  $-10.0\pm 6.9$ &  $10.0\pm6.9$ &  $3.8\pm 7.7$ \\
$-20.0\pm4.6$ &  $-5.4\pm 5.4$ &  $4.6\pm4.6$ &  $9.2\pm 5.4$ \\
$-2.3\pm3.8$ &  $6.2\pm 3.1$ &  $-3.8\pm3.8$ &  $1.5\pm 3.8$ \\
$-16.2\pm6.9$ &  $22.3\pm 6.9$ &  $-3.1\pm6.2$ &  $10.0\pm 6.2$ \\
$-6.2\pm7.7$ &  $18.5\pm 9.2$ &  $6.2\pm9.2$ &  $13.1\pm 7.7$ \\
$3.1\pm3.1$ &  $3.8\pm 3.8$ &  $-10.8\pm3.1$ &  $5.4\pm 3.1$ \\
$-10.0\pm8.5$ &  $1.5\pm 7.7$ &  $-1.5\pm7.7$ &  $3.1\pm 8.5$ \\
$0.8\pm8.5$ &  $6.9\pm 8.5$ &  $-13.1\pm8.5$ &  $2.3\pm 9.2$ \\
$-16.2\pm6.9$ &  $4.6\pm 7.7$ &  $0.8\pm6.9$ &  $26.2\pm 8.5$ \\
$1.5\pm6.2$ &  $-21.5\pm 6.2$ &  $12.3\pm6.2$ &  $-6.2\pm 5.4$ \\
$0.8\pm6.9$ &  $-1.5\pm 7.7$ &  $-7.7\pm7.7$ &  $-13.1\pm 7.7$ \\
$7.7\pm9.2$ &  $-14.6\pm 8.5$ &  $14.6\pm8.5$ &  $3.8\pm 9.2$ \\
$0.0\pm4.6$ &  $6.9\pm 3.8$ &  $2.3\pm3.8$ &  $7.7\pm 3.8$ \\
$-16.2\pm6.9$ &  $-10.0\pm 8.5$ &  $1.5\pm7.7$ &  $11.5\pm 7.7$ \\
$-1.5\pm4.6$ &  $0.8\pm 3.8$ &  $-3.1\pm4.6$ &  $1.5\pm 3.8$ \\
$-5.4\pm10.0$ &  $10.8\pm 12.3$ &  $-20.8\pm10.0$ &  $-14.6\pm 10.8$ \\
\noalign{\smallskip}\hline
\end{tabular}
\end{center}
\end{table}
To compare with the similar D\"usseldorf experiment of
Ref.\cite{schiller}, one should just re-nominate the two sets
\BE(C_0,C_{s1},C_{c1},C_{s2},C_{c2})\to (C_0,C_1,C_2,C_3,C_4)\EE
 \BE (S_0,S_{s1},S_{c1},S_{s2},S_{c2})\to
(B_0,B_1,B_2,B_3,B_4)\EE and introduce an overall factor of two for
the frequency shift since, in this case, two orthogonal cavities are
maintained in a state of active rotation.

As suggested by the same authors of Refs.\cite{peters,schiller}, it
is safer to concentrate on the observed time modulation of the
signal, i.e. on the quantities ${C}_{s1},{C}_{c1},{C}_{s2},{C}_{c2}$
and on their ${S}$-counterparts. In fact, the constant components
${C}_0 $ and $S_0\equiv B_0$ are likely affected by spurious
systematic effects. The experimental $C_k$ and $S_k$ coefficients,
as extracted from Fig. 2 of Ref.\cite{joint}, are reported in our
Tables 1 and 2. The quoted errors are both statistical and
systematical. For the quantities we are considering, according to
Ref.\cite{peters}, this latter component should be very small.

\begin{table}

\caption{The experimental $S_k\equiv B_k$ coefficients as extracted
from Fig. 2 of Ref.\cite{joint}.}
\begin{center}
\label{tab:1}
\begin{tabular}{llll}
\hline\noalign{\smallskip} $S_{s1}[{\rm x}10^{-16}]$ & $S_{c1}[{\rm
x}10^{-16}]$ & $S_{s2}[{\rm x}10^{-16}]$ &
$S_{c2}[{\rm x}10^{-16}]$   \\
\noalign{\smallskip}\hline\noalign{\smallskip} \hline
$2.3\pm3.8$ &  $-10.8\pm 4.6$ &  $-4.6\pm4.6$ &  $-4.6\pm 4.6$ \\
$14.6\pm6.9$ &  $-10.0\pm 6.9$ &  $-6.9\pm6.9$ &  $1.5\pm 6.2$ \\
$-1.5\pm4.6$ &  $-5.8\pm 3.5$ &  $3.8\pm3.8$ &  $-2.3\pm 3.8$ \\
$-8.5\pm3.8$ &  $8.5\pm 3.8$ &  $-10.0\pm3.8$ &  $4.6\pm 4.6$ \\
$3.1\pm7.7$ &  $-3.8\pm 8.5$ &  $-0.8\pm6.9$ &  $-6.9\pm 6.9$ \\
$1.5\pm6.2$ &  $3.1\pm 6.2$ &  $-10.8\pm6.2$ &  $1.5\pm 6.2$ \\
$9.2\pm6.2$ &  $3.8\pm 5.4$ &  $5.4\pm5.4$ &  $-1.5\pm 6.2$ \\
$-3.8\pm8.5$ &  $-2.3\pm 8.5$ &  $-16.2\pm8.5$ &  $-14.6\pm 8.5$ \\
$-4.6\pm6.2$ &  $10.0\pm 6.9$ &  $9.2\pm6.2$ &  $1.5\pm 6.2$ \\
$3.8\pm5.4$ &  $-4.6\pm 4.6$ &  $5.8\pm5.0$ &  $4.6\pm 4.6$ \\
$3.1\pm3.1$ &  $1.5\pm 3.1$ &  $-3.8\pm3.8$ &  $-1.5\pm 3.1$ \\
$3.1\pm7.7$ &  $2.3\pm 6.9$ &  $2.3\pm6.9$ &  $0.8\pm 6.9$ \\
$-3.8\pm3.8$ &  $0.4\pm 5.0$ &  $3.8\pm3.8$ &  $-2.3\pm 3.8$ \\
$5.4\pm3.8$ &  $1.5\pm 3.1$ &  $6.2\pm3.1$ &  $3.1\pm 3.1$ \\
$-14.6\pm5.4$ &  $20.0\pm 6.2$ &  $3.8\pm5.4$ &  $6.5\pm 5.8$ \\
$10.0\pm6.9$ &  $-19.2\pm 8.5$ &  $16.9\pm7.7$ &  $-16.2\pm 6.9$ \\
$-2.3\pm3.8$ &  $2.3\pm 3.8$ &  $8.5\pm3.8$ &  $-8.5\pm 3.8$ \\
$2.3\pm6.9$ &  $-10.0\pm 6.9$ &  $-8.5\pm6.9$ &  $1.5\pm 7.7$ \\
$-10.0\pm8.5$ &  $3.8\pm 8.5$ &  $-5.4\pm8.5$ &  $-3.1\pm 9.2$ \\
$0.0\pm7.7$ &  $6.2\pm 7.7$ &  $10.8\pm7.7$ &  $-4.6\pm 9.2$ \\
$2.3\pm5.4$ &  $-29.2\pm 6.2$ &  $9.2\pm6.2$ &  $-8.5\pm 5.4$ \\
$-6.9\pm6.9$ &  $-0.8\pm 6.9$ &  $-11.5\pm6.9$ &  $-16.2\pm 6.9$ \\
$-6.2\pm9.2$ &  $-7.7\pm 9.2$ &  $10.8\pm9.2$ &  $-4.6\pm 9.2$ \\
$5.4\pm3.8$ &  $8.5\pm 3.8$ &  $-8.5\pm3.8$ &  $0.8\pm 3.8$ \\
$7.7\pm7.7$ &  $-1.5\pm 9.2$ &  $-8.5\pm8.5$ &  $5.4\pm 8.5$ \\
$-0.8\pm3.8$ &  $-0.8\pm 3.8$ &  $-2.3\pm3.8$ &  $3.8\pm 3.8$ \\
$-16.2\pm11.5$ &  $6.2\pm 13.8$ &  $-25.4\pm11.5$ &  $-16.2\pm 11.5$ \\
\noalign{\smallskip}\hline
\end{tabular}
\end{center}
\end{table}


\section{The daily average amplitude of the signal}

For our analysis, we shall re-write Eq.(\ref{basic2}) as follows \BE
\label{basic3}
      {{\Delta \nu (t)}\over{\nu_0}} =
      A(t)\cos (2\omega_{\rm rot}t -2\theta_0(t))
\EE with \BE \label{interms}
C(t)=A(t)\cos2\theta_0(t)~~~~~~~~S(t)=A(t)\sin2\theta_0(t)\EE
$\theta_0(t)$ representing the instantaneous direction of a
hypothetical ether-drift effect in the plane of the interferometer.

Eq.(\ref{basic3}), while fully equivalent to Eq.(\ref{basic2}),
introduces in the analysis a positive-definite quantity, the
amplitude of the signal $A(t)$. The interest in a computation of
$A(t)$ can be easily understood by comparing the two elementary
amplitudes $C(t)$ and $S(t)$ with the cartesian coordinates $x$ and
$y$ of a 2-dimensional particle motion and $A(t)$ with the radial
coordinate $r=\sqrt{x^2+y^2}$, so that $x=r\cos \theta$ and $y=r
\sin \theta$. For definiteness let us denote by $x_i \pm \Delta x_i$
and $y_i \pm \Delta y_i$ the individual measurements of the particle
position and assume that, to a good accuracy, the average values are
$\bar{ x} =\bar{ y}=0$. It goes without saying that this situation
does not determine all properties of the motion since there might be
a plenty of non-trivial rotationally invariant motions that differ
for the average radius. To determine this other parameter, one has
to compute the radial distances $r_i=\sqrt{x^2_i + y^2_i}$ for the
individual measurements and finally take their average. Of course,
since now one deals with the positive-definite quantities $r_i$,
their mean value $ \bar{ r}$ will be definitely different from
$\sqrt{\bar{ x}^2 + \bar{ y}^2 } = 0$ and will depend on the
experimental accuracy of the individual values $r_i \pm \Delta r_i$.
Assuming that the errors for the elementary entries $x_i$ and $y_i$
belong to a normal distribution, the errors $\Delta r_i$ can be
computed according to standard error propagation for composite
observables, namely \BE \Delta r_i =\sqrt{\cos^2\theta_i (\Delta
x_i)^2 + \sin^2\theta_i (\Delta y_i)^2} \EE Therefore, the final
answer about a possible non-zero $ \bar{ r}$ will depend on the
confidence level of the point $r=0$, i.e. on the distributions of
the ratios $R_i=r_i/\Delta r_i$. At the same time, when using the
vectorial observables $S(t)$ and $C(t)$, noise tends to average out
to zero for large sample of data. With  a positive-definite quantity
as $A(t)$ this does not occur. Therefore one has to estimate the
effect of noise that might mimic a true signal. This point will be
discussed in Sect.4 .

After these preliminary considerations, let us return to our basic
amplitude $A(t)$ by observing that, in the framework where the
amplitudes $S(t)$ and $C(t)$ are represented as in Sect.2, it can be
expressed in terms of $v(t)$, the magnitude of the projection of the
cosmic Earth's velocity in the plane of the interferometer, and of
the light anisotropy parameter as
\BE \label{amplitude1}
       A(t)= {{1}\over{2}}|B| {{v^2(t) }\over{c^2}}
\EE
To compute $v(t)$, we shall use the theoretical relations given by
Nassau and Morse \cite{nassau} from which $A(t)$, $S(t)$ and $C(t)$
can be obtained up to an overall proportionality constant. These
expressions are valid for series of short-period observations, as
those performed in Refs.\cite{joint,peters,schiller}, where, within
each experimental session, the kinematical parameters of the cosmic
velocity ${\bf{V}}$ are not appreciably modified by changes in the
Earth's orbital motion around the Sun.

In this approximation, by introducing the magnitude $V$ of the full
Earth's velocity with respect to a hypothetic preferred frame
$\Sigma$, its right ascension $\alpha$ and angular declination
$\gamma$, we get

\BE \label{cosine}
       \cos z(t)= \sin\gamma\sin \phi + \cos\gamma
       \cos\phi \cos(\tau-\alpha)
\EE \BE
       \sin z(t)\cos\theta_0(t)= \sin\gamma\cos \phi -\cos\gamma
       \sin\phi \cos(\tau-\alpha)
\EE \BE
       \sin z(t)\sin\theta_0(t)= \cos\gamma\sin(\tau-\alpha) \EE
\BE \label{projection}
       v(t)=V \sin z(t) ,
\EE
where $z=z(t)$ is the zenithal distance of ${\bf{V}}$, $\phi$ is the
latitude of the laboratory and again $\tau=\omega_{\rm sid}t$
denotes the sidereal time of the observation in degrees. As one can
check, by using the above relations, together with
Eqs.(\ref{interms}) and(\ref{amplitude1}), and replacing
$\chi=90^o-\phi$, one re-obtains the expansions for $C(t)$ and
$S(t)$ reported in Eqs.(\ref{amorse1})-(\ref{CS2}).

After having obtained this first consistency check, we can now
replace Eq.~(\ref{projection}) into Eq.~(\ref{amplitude1}) and, by
adopting a notation of the type in
Eqs.(\ref{amorse1})-(\ref{amorse2}), express $A(t)$ as
\BE \label{amorse}
       A(t) = A_0 +
       A_1\sin\tau +A_2 \cos\tau
        +  A_3\sin(2\tau) +A_4 \cos(2\tau)
\EE
where
\BE \label{aa0}
       A_0 ={{1}\over{2}} |K|
       \left(1- \sin^2\gamma\cos^2\chi
       - {{1}\over{2}} \cos^2\gamma\sin^2\chi \right)
\EE
\BE \label{a1}
       A_1=-{{1}\over{4}}|K| \sin 2\gamma
       \sin\alpha \sin 2\chi
~~~~~~~~~~~~~~~
       A_2=-{{1}\over{4}}|K| \sin 2\gamma
       \cos\alpha \sin 2\chi
\EE
\BE \label{a3}
       A_3=-{{1}\over{4}} |K| \cos^2 \gamma
       \sin 2\alpha \sin^2 \chi
~~~~~~~~~~~~~~~
       A_4=-{{1}\over{4}} |K| \cos^2 \gamma
       \cos 2\alpha \sin^2 \chi
\EE
On the basis of the above relations, it is now possible to extract
the average amplitude $A_0$ from the published data of
Ref.\cite{joint}. On the other hand, no information on the phase of
the signal $\theta_0(t)$ can be obtained. To this end, in fact, one
would need the full amplitudes $S(t)$ and $C(t)$ at the sidereal
times of the observations and these are not available.

To obtain $A_0$, we observe that the daily averaging of the signal
(here denoted by $\langle..\rangle$), when used in Eq.(\ref{amorse})
produces the relation \BE \label{amplitude0}\langle ~A^2(t)~
\rangle= A^2_0+ {{1}\over{2}}(A^2_{1}+A^2_{2}+A^2_{3}+A^2_{4})\EE On
the other hand, using Eqs.(\ref{amorse1}), (\ref{amorse2}) and
(\ref{interms}), one also obtains
 \BE
\label{amplitude}\langle ~A^2(t)~ \rangle= \langle ~C^2(t) +S^2(t)~
\rangle=C^2_0 + S^2_0 +
{{1}\over{2}}(C^2_{11}+S^2_{11}+C^2_{22}+S^2_{22})\EE where we have
introduced the combinations
\BE \label{csid}
      {C}_{11}\equiv \sqrt{{C}^2_{s1}
      + {C}^2_{c1}}
~~~~~~~~~~~~~~~~
      {C}_{22}\equiv \sqrt{{C}^2_{s2}
      + {C}^2_{c2}}
\EE
 \BE \label{s2sid}
      {S}_{11}\equiv \sqrt{{S}^2_{s1}
      + {S}^2_{c1}}
~~~~~~~~~~~~~~~~
 {S}_{22}\equiv \sqrt{{S}^2_{s2}
      + {S}^2_{c2}}
\EE
As one can check, replacing the expressions (\ref{aa0})-(\ref{a3}),
Eq.(\ref{amplitude0}) gives exactly the same result that one would
obtain replacing the values for the C- and S- coefficients in
Eq.(\ref{amplitude}). Therefore, one can combine the two relations
and get \BE \label{final} A^2_0(1+r)= C^2_0 + S^2_0 +
{{1}\over{2}}(C^2_{11}+S^2_{11}+C^2_{22}+S^2_{22})\equiv
C^2_0+S^2_0+Q^2  \EE where\BE \label{Q} Q= \sqrt{
{{1}\over{2}}(C^2_{11}+S^2_{11}+C^2_{22}+S^2_{22})} \EE
 and
 \BE r\equiv
{{1}\over{2A^2_0}}(A^2_{1}+A^2_{2}+A^2_{3}+A^2_{4}) \EE

To evaluate $A_0$ we shall proceed as follows. On the one hand, we
shall compute the ratio $r=r(\gamma,\chi)$ using the theoretical
expressions Eqs.(\ref{aa0})-(\ref{a3}). This gives \BE \label{range}
0\leq r\leq 0.4\EE for the latitude of the laboratories in Berlin
\cite{peters} and D\"usseldorf \cite{schiller} in the full range $0
\leq |\gamma|\leq \pi/2$. On the other hand, we shall adopt the
point of view of the authors of Refs.\cite{joint,peters,schiller}
that, even when large non-zero values of $C_0$ and $S_0$ are
obtained (compare with the value $C_0=(-59.0 \pm 3.4 \pm 3.0)\cdot
10^{-16}$ of Ref.\cite{schiller} and with the large scatter of the
data reported in in Fig.2 of Ref.\cite{joint} or in Fig.3 of
Ref.\cite{peters}), tend to consider these individual determinations
as spurious effects.

\begin{table}
\caption{ The experimental values of Ref.\cite{joint} for the
combinations of $C-$ and $S-$ coefficients defined in
Eqs.(\ref{csid})-(\ref{s2sid}) and the resulting  $Q$ from
Eq.(\ref{Q}). For simplicity, we report symmetrical errors.}
\begin{center}
\label{tab:1}
\begin{tabular}{lllll}
\hline\noalign{\smallskip}
$ {C}_{11}  [{\rm x}10^{-16}]$ & $
{C}_{22}  [{\rm x}10^{-16}]$ & $ {S}_{11}  [{\rm x}10^{-16}]$ &
$ {S}_{22}  [{\rm x}10^{-16}]$ & $ Q [{\rm x}10^{-16}]$  \\
\noalign{\smallskip}\hline\noalign{\smallskip} \hline
$13.6\pm5.0$ &  $1.5\pm 4.6$ &  $11.0\pm4.6$ &  $6.5\pm 4.6$& ${13.3}\pm 3.4$ \\
$5.2\pm6.3$ &  $5.8\pm 6.5$ &  $17.7\pm6.9$ &  $7.1\pm 6.9$& ${14.6} \pm 4.8$ \\
$2.3\pm3.8$ &  $5.2\pm 3.8$ &  $6.0\pm3.5$ &  $4.5\pm 3.8$ & ${6.6}\pm 2.6$ \\
$12.4\pm4.0$ &  $14.6\pm 3.8$ &  $12.0\pm3.8$ &  $11.0\pm 4.0$ &${17.8}\pm 2.8$\\
$12.7\pm8.5$ &  $12.4\pm 8.3$ &  $4.9\pm8.2$ &  $7.0\pm 6.9$ &${14.0}\pm 5.8$\\
$8.1\pm5.9$ &  $7.1\pm 5.4$ &  $3.4\pm6.2$ &  $10.9\pm 6.2$ &${11.1}\pm 4.2 $\\
$12.0\pm6.0$ &  $7.7\pm 5.9$ &  $10.0\pm6.0$ &  $5.6\pm 5.4$ &${13.0} \pm 4.2 $\\
$8.3\pm9.2$ &  $13.2\pm 9.0$ &  $4.5\pm8.5$ &  $21.8\pm 8.5$ &${19.2} \pm 6.1$\\
$8.7\pm7.0$ &  $7.4\pm 6.2$ &  $11.0\pm6.8$ &  $9.4\pm 6.2$ &${13.0}\pm 4.7$\\
$9.4\pm4.6$ &  $10.5\pm 5.4$ &  $6.0\pm4.9$ &  $7.4\pm 4.9$ &${12.0}\pm 3.5$\\
$5.2\pm3.2$ &  $3.1\pm 3.1$ &  $3.4\pm3.1$ &  $4.1\pm 3.7$ &${5.7}\pm 2.4$\\
$17.1\pm7.4$ &  $10.7\pm 7.0$ &  $3.8\pm7.4$ &  $2.4\pm 6.9$ &${14.6}\pm 5.2$\\
$20.7\pm4.7$ &  $10.3\pm 5.2$ &  $3.9\pm3.9$ &  $4.5\pm 3.8$ &${16.9}\pm 3.3 $\\
$6.6\pm3.2$ &  $4.1\pm 3.8$ &  $5.6\pm3.8$ &  $6.9\pm 3.1$ &${8.3} \pm 2.4$\\
$27.5\pm6.9$ &  $10.5\pm 6.2$ &  $24.8\pm5.9$ &  $7.6\pm 5.7$ &${27.7}\pm 4.5$\\
$19.5\pm9.1$ &  $14.5\pm 8.0$ &  $21.7\pm8.2$ &  $23.4\pm 7.3$& ${28.3}\pm 5.7$ \\
$4.9\pm3.6$ &  $12.0\pm 3.1$ &  $3.3\pm3.8$ &  $12.0\pm 3.8$& ${12.7} \pm 2.5$ \\
$10.1\pm8.4$ &  $3.4\pm 8.3$ &  $10.3\pm6.9$ &  $8.6\pm 6.9$ & ${12.1}\pm 5.3$ \\
$7.0\pm8.5$ &  $13.3\pm 8.5$ &  $10.7\pm8.5$ &  $6.2\pm 8.7$ &${13.7}\pm 6.0$\\
$16.8\pm7.0$ &  $26.2\pm 8.5$ &  $6.2\pm7.7$ &  $11.7\pm 8.0$ &${23.9}\pm 5.7$\\
$21.6\pm6.2$ &  $13.8\pm 6.0$ &  $29.3\pm6.1$ &  $12.5\pm 5.8$ &${28.9}\pm 4.3 $\\
$1.7\pm7.5$ &  $15.2\pm 7.7$ &  $7.0\pm6.9$ &  $19.9\pm 6.9$ &${18.4} \pm 5.1 $\\
$16.5\pm8.6$ &  $15.1\pm 8.5$ &  $9.9\pm9.2$ &  $11.7\pm 9.2$ &${19.2} \pm 6.2$\\
$6.9\pm3.8$ &  $8.0\pm 3.8$ &  $10.0\pm3.8$ &  $8.5\pm 3.8$ &${11.9}\pm 2.7$\\
$19.0\pm7.4$ &  $11.6\pm 7.7$ &  $7.8\pm7.8$ &  $10.0\pm 8.5$ &${18.1}\pm 5.4$\\
$1.7\pm4.5$ &  $3.4\pm 4.5$ &  $1.1\pm3.8$ &  $4.5\pm 3.8$ &${4.2}\pm 2.9$\\
$12.0\pm11.9$ &  $25.4\pm 10.3$ &  $17.3\pm11.9$ &  $30.1\pm 11.5$ &${31.6}\pm 7.9$\\
\noalign{\smallskip}\hline
\end{tabular}
\end{center}
\end{table}
 This means to set in Eq.(\ref{final}) \BE S_0
=\langle~ A(t)\sin 2\theta_0(t)~\rangle\sim 0\EE  \BE C_0 =\langle~
A(t)\cos 2\theta_0(t)~\rangle\sim 0\EE We can thus define a daily
average amplitude, say $\hat{A}_0$, which is determined in terms of
$C_{11}$, $S_{11}$, $C_{22}$ and $ S_{22}$ alone as \BE \label{AQ}
\hat{A}_0= {{Q}\over{\sqrt{1+r}}} \sim (0.92 \pm 0.08)Q \EE Here the
uncertainty takes into account the numerical range of $r$ in
Eq.(\ref{range}). This value provides, in any case, a {\it lower
bound} to its true experimental value since \BE \label{lower}
A_0=\sqrt{ {{C^2_0 + S^2_0 +Q^2}\over{1+r}}
 } \geq \hat{A}_0 \EE
The data for the various coefficients are reported in our Table 3
together with the quantity $Q$. The errors on the various
coefficients have been computed, according to standard error
propagation for composite observables, starting from the errors for
the basic entries $C_k$ and $S_k$ reported in our Tables 1 and 2.

\section{The value of $A_0$ and its experimental uncertainty}

In this section we shall now try to deduce the average value of
$A_0$ (or more precisely its lower bound) for the Berlin experiment
\cite{joint,peters}. To this end, a first simple choice could be to
take the weighted average $\bar{Q}$ of the 27 values of Q reported
in Table 3 and then use Eq.(\ref{AQ}). This straightforward strategy
gives \BE \label{Qmean} \bar{Q}=\left(13.0 \pm 0.7 \right)\cdot
10^{-16} \EE

In the above relation the quoted error should be understood as
purely statistical. In fact, by taking the weighted average of a
large number N of measurements, one obtains an error on the mean
that vanishes as $1/\sqrt{N}$ in the limit $N\to \infty$. To
estimate a possible systematic error to be added in Eq.(\ref{Qmean})
we shall make the reasonable assumption that systematic
uncertainties affect the value $Q_i$ of the i-th experimental
session by an amount $Q^{sys}_i$ which is comparable to $\Delta
Q_i$. To get a numerical value, we shall follow three different
methods that turn out to give similar results.

In a first approach, one could estimate the systematic error on the
mean by simply averaging the 27 determinations $Q^{sys}_i\sim \Delta
Q_i$ reported in Table 4. This gives the value \BE
\label{noise1}\overline{\Delta Q}^{(1)}\sim {{1}\over{27}}
 \sum_i ~Q^{sys}_i\sim 4.4\cdot 10^{-16} \EE
As a second approach, one can estimate a mean systematic component
by using the weighted average expression \BE \overline {\Delta
Q}^{(2)} \sim \sum_i {{Q^{sys}_i}\over{(\Delta Q_i)^2}}~~
\left(\sum_i {{1}\over{(\Delta Q_i)^2}}\right)^{-1} \EE In this
case, by using again the data reported in Table 4, we obtain \BE
\label{noise2} \overline{\Delta Q}^{(2)}\sim 3.5\cdot 10^{-16} \EE
Finally, one can also consider the values of the signal and estimate
the noise for the i-th experimental session from the ratio
$R_i=Q_i/\Delta Q_i$. Taking the arithmetic mean of the 27
determinations $R_i$ reported in Table 4 gives a value $\bar{R}\sim
3.6$ that, when using the weighted average $\bar{Q}=13.0\cdot
10^{-16}$, gives the other estimate \BE \label{noise3}
\overline{\Delta Q}^{(3)}\sim {{\bar{Q}}\over{ \bar{R}}} \sim
3.6\cdot 10^{-16} \EE Therefore, by averaging the three different
determinations in Eqs.(\ref{noise1}), (\ref{noise2}) and
(\ref{noise3}), one obtains the value \BE \overline{\Delta Q}\sim
3.8\cdot 10^{-16}\EE that we shall take as our final estimate of the
systematic error on the mean thus replacing Eq.(\ref{Qmean}) with
 \BE \label{Qmean1} \bar{Q}=\left(13.0 \pm 0.7 \pm 3.8\right)\cdot
10^{-16} \EE From this, by using Eq.(\ref{AQ}), we get \BE
\label{final1} \hat{A}_0=(12.0 \pm 1.0 \pm 3.5)\cdot
10^{-16}~~~~~~~~~~~~{\rm Ref.\cite{joint}} \EE where the former
error is due to the variation $0 \leq r \leq 0.4$  and the latter
reflects our estimate of the noise.

As a further control of the validity of our Eq.(\ref{Qmean1}), we
report in Tables 5-7 the $C$ and $S$ coefficients of
Ref.\cite{schiller}. As one can check the value $Q=(13.1 \pm
2.1)\cdot 10^{-16}$ is in excellent agreement with our result
(\ref{Qmean1}) and corresponds to \BE \label{mean2} \hat{A}_0= (12.1
\pm 1.0 \pm 2.1)\cdot 10^{-16}~~~~~~~~~~~~~~~{\rm
Ref.\cite{schiller}}\EE in perfect agreement with (\ref{final1}).

\begin{table}
\caption{For each experimental session, we report the values $Q_i$,
$\Delta Q_i$ and the ratio $R_i=Q_i/\Delta Q_i$ from Table 3.}
\begin{center}
\label{tab:1}
\begin{tabular}{lll}
\hline\noalign{\smallskip}
 $ Q_i  [{\rm x}10^{-16}]$ & $ \Delta Q_i  [{\rm
x}10^{-16}]$ & $R_i= Q_i/\Delta Q_i  $  \\
\noalign{\smallskip}\hline\noalign{\smallskip} \hline
$13.3$ & ~~~ $3.4$ &~~~  $3.9$  \\
$14.6$ & ~~~ $4.8$ & ~~~ $3.0$  \\
$6.6$ &  ~~~ $2.6$ & ~~~ $2.5$  \\
$17.8$ & ~~~ $2.8$ & ~~~ $6.3$ \\
$14.0$ & ~~~ $5.8$ & ~~~ $2.5$ \\
$11.1$ & ~~~ $4.2$ & ~~~ $2.6$ \\
$13.0$ & ~~~ $4.2$ & ~~~ $3.1$ \\
$19.2$ & ~~~ $6.1$ & ~~~ $3.1$ \\
$13.0$ & ~~~ $4.7$ & ~~~ $2.8$ \\
$12.0$ & ~~~ $3.5$ & ~~~ $3.4$ \\
$5.7$ &  ~~~ $2.4$ &  ~~~ $2.4$ \\
$14.6$ &  ~~~ $5.2$ & ~~~ $2.8$ \\
$16.9$ &  ~~~ $3.3$ & ~~~ $5.1$ \\
$8.3$ &  ~~~ $2.4$ &  ~~~ $3.4$ \\
$27.7$ & ~~~ $4.5$ & ~~~ $6.2$  \\
$28.3$ & ~~~ $5.7$ & ~~~ $5.0$  \\
$12.7$ & ~~~ $2.5$ & ~~~ $5.1$ \\
$12.1$ & ~~~ $5.3$ & ~~~ $2.3$ \\
$13.7$ & ~~~ $6.0$ & ~~~ $2.3$ \\
$23.9$ & ~~~ $5.7$ & ~~~ $4.2$ \\
$28.9$ & ~~~ $4.3$ & ~~~ $6.7$ \\
$18.4$ & ~~~ $5.1$ & ~~~ $3.6$ \\
$19.2$ & ~~~ $6.2$ & ~~~ $3.1$ \\
$11.9$ & ~~~ $2.7$ & ~~~ $4.4$ \\
$18.1$ & ~~~ $5.4$ & ~~~ $3.3$ \\
$4.2$ & ~~~  $2.9$ &  ~~~ $1.4$ \\
$31.6$ &~~~  $7.9$ & ~~~ $4.0$ \\
\noalign{\smallskip}\hline
\end{tabular}
\end{center}
\end{table}

\begin{table}
\caption{The experimental $C-$coefficients as reported in
Ref.\cite{schiller}. }
\begin{center}
\label{tab:1}
\begin{tabular}{llll}
\hline\noalign{\smallskip}
$C_{s1}[{\rm x}10^{-16}]$ & $C_{c1}[{\rm
x}10^{-16}]$ & $C_{s2}[{\rm x}10^{-16}]$ &
$C_{c2}[{\rm x}10^{-16}]$   \\
\noalign{\smallskip}\hline\noalign{\smallskip} \hline
$-3.0\pm 2.0$ &  $11.0\pm 2.5$ &  $1.0\pm2.5$ &  $0.1\pm 2.5$ \\
\noalign{\smallskip}\hline
\end{tabular}
\end{center}
\end{table}

\begin{table}
\caption{The $S-$coefficients of Ref.\cite{schiller}. The values for
the S-coefficients, constrained by the authors of
Ref.\cite{schiller} in their fits to the data to the theoretical
predictions Eqs.(\ref{s1}) and (\ref{s2}), have been deduced from
Table 5 using these relations. }
\begin{center}
\label{tab:1}
\begin{tabular}{llll}
\hline\noalign{\smallskip} $S_{s1}[{\rm x}10^{-16}]$ & $S_{c1}[{\rm
x}10^{-16}]$ & $S_{s2}[{\rm x}10^{-16}]$ &
$S_{c2}[{\rm x}10^{-16}]$   \\
\noalign{\smallskip}\hline\noalign{\smallskip} \hline
$-14.1\pm 3.2$ &  $-3.8\pm 2.6$ &  $-0.1\pm2.5$ &  $1.0\pm 2.5$ \\
\noalign{\smallskip}\hline
\end{tabular}
\end{center}
\end{table}

\begin{table}
\caption{ The values of Ref.\cite{schiller} for the combinations of
$C-$ and $S-$ coefficients defined in Eqs.(\ref{csid})-(\ref{s2sid})
and the resulting $Q$ defined in Eq.(\ref{Q}), as computed from
Tables 5 and 6. For simplicity, we report symmetrical errors.}
\begin{center}
\label{tab:1}
\begin{tabular}{lllll}
\hline\noalign{\smallskip}
 $ {C}_{11}  [{\rm x}10^{-16}]$ & $
{C}_{22}  [{\rm x}10^{-16}]$ & $ {S}_{11}  [{\rm x}10^{-16}]$ &
$ {S}_{22}  [{\rm x}10^{-16}]$& $ Q [{\rm x}10^{-16}]$ \\
\noalign{\smallskip}\hline\noalign{\smallskip} \hline
$11.4\pm 2.5 $ &  $1.0\pm 2.5$ &  $14.6\pm3.3$ &  $1.0\pm 2.5$ &${13.1}\pm 2.1$\\
\noalign{\smallskip}\hline
\end{tabular}
\end{center}
\end{table}

\section{An effective refractive index for the vacuum}

In this section, we shall point out that the two experimental values
in Eqs.(\ref{final1}) and (\ref{mean2}) are well consistent with the
theoretical prediction \BE \label{theory1} A^{\rm th}_0 =
{{1}\over{2}} |B_{\rm th}| {{v^2_0}\over{c^2}} \sim (9.7 \pm
3.5)\cdot 10^{-16} \EE of Refs.\cite{pla,cimento}. This was
obtained, in connection with the anisotropy parameter \cite{pagano}
$|B_{\rm th}|\sim 42\cdot 10^{-10}$, after inserting the average
cosmic velocity (projected in the plane of the interferometer)
$v_0=(204 \pm 36)$ km/s that derives from a re-analysis
\cite{pla,cimento} of the classical ether-drift experiments. Due to
this rather large theoretical uncertainty, the different locations
of the various laboratories and any other kinematical property of
the cosmic motion can be neglected in a first approximation.

For a proper comparison, we also remind that in
Refs.\cite{pla,cimento} the frequency shift was parameterized as \BE
{{\Delta\nu(\theta)}\over{\nu_0}}= |B_{\rm th}| {{v^2}\over{c^2}}
\cos 2\theta\EE This relation is appropriate for a symmetrical
apparatus with two rotating orthogonal lasers, where the difference
$\bar{c}(\theta)-\bar{c}(0)$ gets replaced by
$\bar{c}(\pi/2+\theta)-\bar{c}(\theta)$ as in the D\"usseldorf
experiment \cite{schiller}, and gives an average amplitude \BE
A^{\rm symm}_0=2A^{\rm th}_0 \sim (19 \pm 7)\cdot 10^{-16}\EE

The theoretical prediction for the anisotropy parameter was obtained
starting from the many analogies that one can establish, in the
weak-field limit, between General Relativity  and a flat-space
description where gravity leads to re-defined masses, space-time
units and an effective vacuum refractive index. This alternative
approach, see for instance Wilson \cite{wilson}, Gordon
\cite{gordon}, Rosen \cite{rosen}, Dicke \cite{dicke}, Puthoff
\cite{puthoff} and even Einstein himself \cite{pre}, before his
formulation of a metric theory of gravity, in spite of the deep
conceptual differences, produces an equivalent description of the
phenomena in a weak gravitational field.

The substantial phenomenological equivalence of the two approaches
was well summarized by Atkinson as follows \cite{atkinson} : "It is
possible, on the one hand, to postulate that the velocity of light
is a universal constant, to define {\it natural} clocks and
measuring rods as the standards by which space and time are to be
judged and then to discover from measurement that space-time is {\it
really} non-Euclidean. Alternatively, one can {\it define} space as
Euclidean and time as the same everywhere, and discover (from
exactly the same measurements) how the velocity of light and natural
clocks, rods and particle inertias {\it really} behave in the
neighborhood of large masses."

This type of analogy, which is preserved by the weak-field classical
tests, is interesting in itself and deserves to be explored. In
fact, "...it is not unreasonable to wonder whether it may not be
better to give up the geometric approach to gravitation for the sake
of obtaining a more uniform treatment for all the various fields of
force that are found in nature" \cite{rosen}.

To distinguish between the two interpretations, one can start from
the zeroth order approximation to the problem embodied in the
Equivalence Principle. As it is well known, this basic principle,
introduced before General Relativity, does not necessarily rely on
the notion of a curved space-time (see e.g. Ref.\cite{kirkwood} or
even the critical point of view of Ref.\cite{synge}). According to
it, for an observer placed in a freely falling frame, local Lorentz
invariance is valid to first order. Therefore, given two space-time
events that differ by $(dx,dy,dz,dt)$, and the space-time metric \BE
ds^2=c^2dt^2- (dx^2+dy^2+dz^2)\EE one gets from $ds^2=0$ the same
speed of light that one would get in the absence of any
gravitational effect.

However, to a closer look, an observer placed on the Earth's surface
is equivalent to a freely-falling frame {\it up to the presence of
the Earth's gravitational field}. In this case, both General
Relativity and the flat-space approach predict the weak-field,
isotropic form of the metric \BE \label{iso} ds^2=
c^2dt^2(1-{{2GM}\over{c^2R}})-(1+{{2GM}\over{c^2R}})
(dx^2+dy^2+dz^2)=c^2d\tau^2-dl^2\EE
 $G$ being Newton's constant and $M$ and $R$ the Earth's mass and
radius. Here $d\tau$ and $dl$ denote respectively the elements of
"proper" time and "proper" length in terms of which, in General
Relativity, one would again deduce from $ds^2=0$ the same universal
value ${{dl}\over{d\tau}}=c$.

On the other hand, in the flat-space approach the condition $ds^2=0$
is interpreted in terms of an effective refractive index for the
vacuum \BE{\cal N}_{\rm vacuum}- 1 \sim {{2GM}\over{c^2R}}\sim
14\cdot 10^{-10}\EE as if Euclidean space would be filled by a very
rarefied "gravitational medium" (let us recall that a moving
dielectric medium acts on light as an effective gravitational field,
see for instance Refs.\cite{gordon,leonard}).

Now, since in the flat-space approach light can be seen isotropic by
only one inertial frame \cite{volkov}, say $\Sigma$, the ether-drift
experiments can clarify whether $\Sigma$ coincides with the Earth's
frame or with a hypothetical preferred frame. In the former case,
corresponding to no anisotropy of the two-way speed of light in the
vacuum, the equivalence between General Relativity and the
gravitational-medium picture would persist. In the latter case, for
the observer sitting on the Earth, there would be off-diagonal
elements in the metric \BE g_{0i} \sim ({\cal N}^2_{\rm vacuum}-
1)~{{V_i}\over{c}} \EE and an anisotropy of the speed of light. Its
value can be estimated directly in flat space, by Lorentz
transforming from the isotropic value $c/{\cal N}_{\rm vacuum}$ in
$\Sigma$, and one predicts an anisotropy of the two-way speed of
light \BE {{\bar{c}(\theta)-\bar{c}(0)}\over{c}}\sim B_{\rm th}
{{V^2}\over{c^2}}\sin^2\theta \EE governed by the parameter
\cite{pagano,pla,cimento} \BE \label{theory2}|B_{\rm th}|\sim
3({\cal N}_{\rm vacuum}- 1 )\sim 42\cdot 10^{-10}\EE   For this
reason, assuming a typical value $V^2/c^2 \sim 10^{-6}$, the present
ether-drift experiments, with their ${\cal O}(10^{-16})$ accuracy,
represent precision probes of the vacuum and of its space-time
structure. In this respect, it is interesting that, independently of
our re-analysis of the data, a $2-3$ $\sigma$ non-zero signal in the
photon sector $\tilde{\kappa}^{XZ}_{e^-}=( -10.3 \pm 3.9) \cdot
10^{-16}$ (see Table II of Ref.\cite{joint}) was reported by the
authors of Ref.\cite{joint} by using the SME parametrization
\cite{sme}.

\section{Summary and conclusions}
In this paper, we have presented a re-analysis of the Berlin
\cite{joint,peters} and D\"usseldorf \cite{schiller} ether-drift
experiments that, by employing rotating optical resonators, attempt
to establish the isotropy of the speed of light in the vacuum to a
level of accuracy ${\cal O}(10^{-16})$. In our re-analysis we have
extracted from the published data a basic observable of an
ether-drift experiment, namely the daily average for the amplitude
of the signal. By denoting with $\langle..\rangle$ the daily average
of any given quantity, this is defined as $A_0=\langle A(t)\rangle$
after re-writing the frequency shift \BE
      {{\Delta \nu (t)}\over{\nu_0}} =
      {S}(t)\sin 2\omega_{\rm rot}t +
      {C}(t)\cos 2\omega_{\rm rot}t
\EE in the form  \BE
      {{\Delta \nu (t)}\over{\nu_0}} =
      A(t)\cos (2\omega_{\rm rot}t -2\theta_0(t))
\EE By further assuming, as the authors of
Refs.\cite{peters,schiller} do, that experimental results providing
large non-zero values for either $\langle C(t) \rangle= C_0$ or
$\langle S(t)\rangle =S_0$ in Eqs.(\ref{amorse1}) and
(\ref{amorse2}) should be interpreted as spurious effects (e.g. due
to thermal drift, non-uniformity of the rotating cavity speed,
misalignment of the cavity rotation axis,...), $A_0$ can be
approximated by its lower bound \BE \hat{A}_0\sim (0.92\pm 0.08)Q\EE
where \BE Q= \sqrt{
{{1}\over{2}}(C^2_{11}+S^2_{11}+C^2_{22}+S^2_{22})} \EE is given in
terms of the coefficients $C_{11},C_{22}, S_{11}, S_{22}$ defined in
Eqs.(\ref{csid})-(\ref{s2sid}). These coefficients, that reflect the
time modulation of the signal, are much less affected by spurious
effects (as compared to $C_0$ and $S_0$) and so should be the value
of $\hat{A}_0$.

Now, the two resulting experimental determinations in
Eqs.(\ref{final1}) and (\ref{mean2}), namely $\hat{A}_0= (12.0 \pm
1.0\pm 3.5)\cdot 10^{-16}$ from Ref.\cite{joint} and $\hat{A}_0=
(12.1\pm 1.0\pm 2.1)\cdot 10^{-16}$ from Ref.\cite{schiller},
besides being in excellent agreement with each other (and with the
previous lower-statistics re-analysis of the Berlin experiment
reported in \cite{previous}) also agree with the theoretical
prediction $(9.7\pm 3.5)\cdot 10^{-16} $ of Refs.\cite{pla,cimento}.
This prediction was obtained in a flat-space description of gravity,
by assuming an average speed with respect to a hypothetical
preferred frame of the type suggested by a re-analysis of the
classical ether-drift experiments. Therefore this non-trivial level
of consistency, between different experiments and with a theoretical
prediction formulated {\it before} the experiments were performed,
supports the conclusion that a non-zero anisotropy of the speed of
light in the vacuum has actually been measured with values of the
anisotropy parameter that are {\it one order of magnitude larger}
than those quoted by the authors of Refs.\cite{peters,schiller}.

A first obvious possibility for this discrepancy is that in our
average we might have grossly underestimated the effect of the
residual noise, even in the {\it time-dependent} part of the signal.
In this case, our error in Eq.(\ref{Qmean1}) would be too small. For
instance, by insisting on a zero signal and deciding that a
vanishing amplitude requires consistency with the value $Q= 0$ to
some definite confidence level, say $10\%$ one side, one might argue
that the correct final estimate for the Berlin experiment should be
$\bar Q \sim (13 \pm 10) \cdot 10^{-16}$. To this end, however, the
individual determinations in Table 4 should exhibit a typical value
$R_i= Q_i/\Delta Q_i \sim 1.3$ which is quite different from their
average result $R_i\sim 3.6$. Analogous considerations apply if the
confidence level were pushed to $5\%$, again one side. In this case,
the final estimate for the Berlin experiment should be $\bar Q \sim
(13 \pm 8) \cdot 10^{-16}$ with individual determinations $R_i=
Q_i/\Delta Q_i \sim 1.6$ in Table 4.

Now, the  individual central values $Q_i$ and their errors $\Delta
Q_i$ reported in Table 4 were computed, starting from the values of
the elementary entries $C_k$ and $S_k$ reported in Tables 1 and 2,
according to the standard statistical rules for composite
observables. Errors such as those needed to obtain $\bar Q \sim (13
\pm 10) \cdot 10^{-16}$ or $\bar Q \sim (13 \pm 8) \cdot 10^{-16}$
would be respectively five or four times larger than that extracted
from the data of Ref.\cite{schiller} after a single experimental
session (see Table 7). This would be equivalent to deny the assumed
high precision of the Berlin experiment.

On the other hand, by accepting our estimate of the errors, the
reasons for the discrepancy with the anisotropy parameter reported
in Refs.\cite{peters,schiller} have to be searched elsewhere. To
this end, we observe preliminarily that introducing $C_{11}$,
$C_{22}$, $S_{11}$ and $S_{22}$ represents the simplest way to
obtain {\it rotationally invariant} combinations, out of the
elementary coefficients $C_{s1}$, $C_{c1}$.. and of their
S-counterparts. Their use can eliminate possible spurious effects,
depending on the relative phases, that enter the delicate splitting
of the signal in its various Fourier components (see the discussion
given by the authors of Ref.\cite{peters} and their note [13]).

Now, suppose that for some reason, during the i-th experimental
session the coefficients $C_{s1}$ and $C_{c1}$ were erroneously
rotated by an angle $\varphi_i$ with respect to their true value.
This means erroneous values \BE
C_{s1}(i)=C_{s1}\cos\varphi_i-C_{c1}\sin\varphi_i \EE and \BE
C_{c1}(i)=C_{c1}\cos\varphi_i+C_{s1}\sin\varphi_i \EE  with
analogous effects for the other parameter pairs $(C_{s2},C_{c2})$,
$(S_{s1},S_{c1})$ and $(S_{c2},S_{s2})$ and an overall shift in the
value of the right ascension $\alpha \to \alpha +\varphi_i$.

Nevertheless, even if this happens, our rotationally invariant
combinations would remain unchanged, i.e. \BE
{C}_{11}(i)=\sqrt{C^2_{s1}(i)+ C^2_{c1}(i)}=\sqrt{C^2_{s1}+
C^2_{c1}}={C}_{11}\EE with similar results for $C_{22}$, $S_{11}$
and $S_{22}$.  Therefore, even substantial levels of phase error,
that can produce zero intersession averages for the various $C_k$
and $S_k$ vectorial coefficients (and thus a vanishing anisotropy
parameter from these averages), would not change our determination
of $\hat{A}_0$. In this sense, $\hat{A}_0$ represents a robust
indicator. Analogous considerations apply to the separation of the
signal in its $S(t)$ and $C(t)$ parts which is obtained by
extracting the components proportional to $\sin 2\omega_{\rm rot}t$
and $\cos 2\omega_{\rm rot}t$. Again, the combination $A^2=S^2+C^2$
is not modified by replacing $\omega_{\rm rot}t \to \omega_{\rm
rot}t + \Omega_i$. This type of argument would also explain the
remarkable agreement between our final estimate Eq.(\ref{Qmean1})
from Ref.\cite{joint} and the value $Q=(13.1 \pm 2.1)\cdot 10^{-16}$
from Ref.\cite{schiller}, in spite of the completely different $C_k$
and $S_k$ coefficients obtained by the two experiments.

Clearly, the simplest way to check our result would be that the
authors of Refs.\cite{joint,peters,schiller} could repeat their
analysis of the data, replacing Eq.(\ref{basic2}) with
Eq.(\ref{basic3}), and compute from scratch $A_0$ (and its lower
bound $\hat{A }_0$) for the various experimental sessions. This
computation, that would only require the elementary algebraic
relations used in this paper, could also provide a powerful
consistency check of the whole experiment. At the present, since
this has not been done, by accepting the errors of
Refs.\cite{joint,peters,schiller} for the elementary entries
$C_{s1}$, $C_{c1}$... and their S-counterparts, our values of $A_0$
are the only existing estimate of this basic physical quantity.

For this reason, since with our analysis one would be driven to
deduce non-trivial consequences, such as the existence of a
preferred frame and a flat-space description of gravity, we
emphasize the importance of comparing different points of view and
approaches to the data to finally achieve a full understanding of
the underlying fundamental physical problem.

\end{document}